\documentclass[preprint,superscriptaddress]{revtex4-2}

\usepackage{graphicx,epstopdf,amsmath,amssymb,amsbsy,amstext} 

\begin{document}

\title {Dynamics of Social Balance on Networks: The Emergence of Multipolar Societies}

\author{Pouya Manshour}
\email{manshour@pgu.ac.ir}
\affiliation{Physics Department, Persian Gulf University, Bushehr 75169, Iran}
\affiliation{Department of Complex Systems, Institute of Computer Science of the Czech Academy of Sciences, Pod Vod\'arenskou v\v{e}\v{z}\'{\i}~2, 182~07~Prague~8, Czech~Republic}
\author{Afshin Montakhab}
\affiliation{Physics Department, Shiraz University, Shiraz 71454, Iran}

\begin{abstract}
Within the context of social balance theory, much attention has
been paid to the attainment and stability of unipolar or bipolar
societies. However, multipolar societies are commonplace in the
real world, despite the fact that the mechanism of their emergence
is much less explored. Here, we investigate the evolution of a
society of interacting agents with friendly (positive)  and enmity
(negative) relations into a final stable multipolar state. Triads
are assigned energy according to the degree of tension they impose
on the network. Agents update their connections in order to
decrease the total energy (tension) of the system, on average. Our
approach is to consider a variable energy $\epsilon\in[0,1]$ for
triads which are entirely made of negative relations.  We show
that the final state of the system depends on the initial density
of the friendly links $\rho_0$.  For initial densities greater
than an $\epsilon$ dependent threshold $\rho^c_0(\epsilon)$
unipolar (paradise) state is reached. However, for $\rho_0 \leq
\rho^c_0(\epsilon)$ multi-polar and bipolar states can emerge. We
observe that the number of stable final poles increases with
decreasing $\epsilon$ where the first transition from bipolar to
multipolar society occurs at $\epsilon^*\approx 0.67$. We end the
paper by providing a mean-field calculation that provides an
estimate for the critical ($\epsilon$ dependent) initial positive
link density, which is consistent with our simulations.
\end{abstract}
\pacs{89.65.-s, 89.75.Hc, 05.40.-a}

\maketitle

\section{ Introduction }
\label{intro} Societies experience unipolar, bipolar and
multipolar phases over time
\cite{deutsch1964multipolar,waltz2010theory}. A
pole can be considered as a sub-community of friendly individuals
that cooperate with each other or are in the same opinion on
some issue. The social polarization is a key concept in sociology,
and as a collective phenomenon, it emerges from complex
interactions among individuals due to income inequality,
economical or political thoughts, globalization, migration,
ethno-cultural diversity, modern communication technologies, and
the integration of states into trans-national entities, such as
the European Union \cite{Caves2004,Schiefer2016}. But, how do such
stable polarized phases arises from rearrangement of local social
interactions? In a series of seminal works, it was assumed that
avoiding distress and conflict is the natural mechanism of
creating such a stability
\cite{Heider:1946,Cartwright:1956aa,harary1953notion}. Although polarization is a common phenomenon in socio-politico-economic settings, the number of competing poles is also an important relevant issue, For example, in the realm of politics, United States is dominated by two major political parties while Italy, on the other hand, has many equally strong political parties. Building consensus and coalitions is crucial in multi-polar society, while unilateral action is a possiblity in a bipolar society.

One of the basic concepts in sociology is the structural balance
which is based on the observational intuition that in society
dynamics, triadic interactions are more fundamental than the
pairwise ones. In this respect, Heider's theory, known as the
balance theory considers the relationship between three elements
includes Person (P), and Other person (O) with an object (X),
known as the POX pattern \cite{Heider:1946}. Heider postulated
that the POX is ``balance" if P and O are friends, and they agree
in their opinion of X. In an unbalanced triad, to reduce the
stress and reach some sort of stability, the individuals alter
their opinions so that the triad becomes balanced. Empirical
examples of Heider’s balance theory have been found in human and
other animal societies
\cite{Harary1961,Doreian1996,szell2010multi,Szell2010,Facchetti2011,Ilany2013}.
Cartwright and Harary demonstrated that a society with two
possible interactions between their individuals can be viewed as a
signed graph with positive (agree) and negative (disagree) links
\cite{Cartwright:1956aa,harary1953notion}. They found that the
society is balanced, if and only if it can be decomposed into two
fully positive-link poles that are joined by negative links, i.e.,
a bipolar state. Dynamical evolution of how such stable states can
reach from an initially unbalanced ones is another important
aspects of research studies
\cite{antal2006social,antal2005dynamics,kulakowski2005heider,radicchi2007social,marvel2011continuous,traag2013dynamical,kulakowski2007some,hummon2003some,altafini2012dynamics,Shojaei2019}.
In such dynamical models, the individuals rearrange their
connections in order to reduce the local or global stress in the
society, for example, continuous-valued links models
\cite{kulakowski2005heider,marvel2011continuous}, balance theory
in asymmetric networks \cite{traag2013dynamical}, disease
spreading on sign networks \cite{Saeedian2017}, memory effects on
the evolution of the links \cite{hassanibesheli2017glassy}, and
phase transition in societies with stochastic individual behaviors
\cite{Shojaei2019,ManshourPRE_2021}, to name a few.

Antal \textit{et. al.} proposed a dynamical model, called
\textit{Constrained Triad Dynamics} (CTD)
\cite{antal2006social,antal2005dynamics}. In CTD, a triad with odd
number of positive links is balanced. If $\Delta_k$ represents a
triad of type $k$ which consists of $k$ negative links; then
triads of $\Delta_0$ and $\Delta_2$ are balanced, while triads of
$\Delta_1$ and $\Delta_3$ are unbalanced. They assumed that the
total number of unbalanced triads $N_{unb}$ cannot increase in an
update event. In each update step, a randomly chosen link changes
its sign, if $N_{unb}$ decreases. If $N_{unb}$ remains constant,
then the chosen link changes its sign with probability $1/2$, and
otherwise, sign of the chosen link does not change. Thus, in each
time step, the system goes into a state that is more balanced than
the previous state, and the system eventually approaches into a
final bipolar state. Indeed, for $\rho_0< 0.65$, where $\rho_0$ is
the initial density of the positive links the society divides into
two equal-size poles and for $\rho_0\geq 0.65$, one pole becomes
dominant and we have a unipolar society (paradise). However, a possible
outcome of CTD dynamics is a jammed state, where the system is
trapped into an unbalanced state, forever. They showed that in
spite of the higher number of such states in comparison with the
balanced ones, the probability of reaching a jammed state vanishes
for large systems. By introducing an energy landscape, the
properties of such jammed states have been studied, extensively
\cite{marvel2009energy,facchetti2012exploring}. Shojaei
\textit{et. al.} proposed in \cite{Shojaei2019} a natural
mechanism to escape from such states by introducing a dynamical
model with an intrinsic randomness, similar to Glauber dynamics in
statistical mechanics \cite{Glauber1963}. They also showed that in
finite networks, the system approaches into a balanced state, if
the randomness is lower than a critical value.

The structural balance theory, applied in all above mentioned
models, implies that individuals always tend to polarize into at
most two communities. This is due to the way that unbalanced
triads are defined, i.e., all triadic relationships with odd
number of negative links ($\Delta_1$ and $\Delta_3$) are
considered to be unbalanced. Such conditions for
balanced/unbalanced triads assert that a friend of my friend or an
enemy of my enemy is my friend, and vice versa. However, it has
been observed in social and political societies that the two types
of unbalanced triads of $\Delta_1$ and $\Delta_3$ are not equally
unbalanced and have also a different incidence rate, i.e.,
$\Delta_3$ triads are more frequent than $\Delta_1$
\cite{szell2010multirelational,belaza2017statistical}. On the
other hand, in order to reach multipolar states, we need to have
triads of type $\Delta_3$ survived in the final state of the
dynamics. In 1967, Davis introduced the clustering theory
\cite{davis1967clustering} which generalizes social balance theory
by stating that in many situations an enemy of one's enemy can
indeed act as an enemy. This means that only triads with two
positive links ($\Delta_1$) are unlikely in real stable networks
and all other types of triads ($\Delta_0$, $\Delta_2$ and
$\Delta_3$) can be present. This is indeed in agreement with
empirical studies in human social networks
\cite{leskovec2010signed,van2011micro}. This form of structural
stability is called weak structural balance, in comparison with
the (strong) structural balance theory defined by Heider
\cite{Heider:1946}.

The dynamical models result in the unipolarity or bipolarity have
been studied extensively, however, the notion of multipolarity are
greatly unexplored in the literature. In this article, by
including the stochasticity of individual's behavior similar to
our previous work \cite{Shojaei2019}, we study the evolution of a
society with interacting individuals, seeking to reduce the
tension in the system, based on an energy minimization formalism.
Accordingly, we include the role of triads of type $\Delta_3$ in
the system dynamics by assigning different energy $\epsilon\in
[0,1]$ to them. We observe that the system quickly approaches into
a final stable balanced state. The final fate of the system can be
either a unipolar, a bipolar or a multipolar state based on
different values of energy $\epsilon$ and initial link density
$\rho_0$. We find that the system transitions from a unipolar
state into a multi-bipolar one when the initial positive link
density $\rho_0$ crosses a critical value $\rho_0^c$ from above.
Indeed, the system approaches a unipolar state for any arbitrary
values of $\epsilon$ when $\rho_0>\rho_0^c$. On the other hand,
when $\rho_0\leq\rho_0^c$ the system reaches a multi-bipolar
state, in which the number of poles increases as $\epsilon$ decreases from the value of $\epsilon^*\approx 0.67$. We end the paper by providing a mean-field calculation for our model which provides a bifurcation diagram and is in line with our numerical simulations.

\section{Model definition}
\label{method}

We consider a network of size $N$, and use a symmetric adjacency matrix $A$, such that $A_{ij}=\pm1$. The positive sign represents friendship, and the negative one represents enmity between two arbitrary nodes $i$ and $j$. For simplicity, we assume that everyone knows everyone else, i.e.,
the dynamics occurs on a fully connected graph, which is appropriate for small real-world networks. For simplicity and without loss of generality, we assign energies $\{u_0,u_1,u_2,u_3\}=\{0,1,0,\epsilon\}$ to triads of type $\{\Delta_0,\Delta_1,\Delta_2,\Delta_3\}$, respectively, where $\epsilon\in [0,1]$. This means that triads $\Delta_0$ and $\Delta_2$ have the minimum possible energy corresponding to their minimum tension they impose on the system and triad $\Delta_1$ has the maximum possible energy which indicates its maximum tension. Triads of type $\Delta_3$ can have any energies in the range of $0$ to $1$, which implies that they can have different degrees of tension based on different values of $\epsilon$. By this definition, we take into account the role of triads of type $\Delta_3$ in the system dynamics, which is in line with empirical observations \cite{szell2010multirelational,belaza2017statistical}. We note here that this model is indeed a generalization of the special case of $\epsilon=1$ that has been studied extensively in our previous work \cite{Shojaei2019}. The total energy of the system is defined as:
\begin{equation}
U=\sum_{i} u^i_{\Delta}/N_{tri}
\label{U}
\end{equation}
where the sum is over all triads and $u_{\Delta}\in \{u_0,u_1,u_2,u_3\}$ and the normalization factor of $N_{tri}=N(N-1)(N-2)/6$ is the total number of triads in the system. It is also appropriate to work with quantity $n_i$ which is the density of triads of type $\Delta_i$, i.e., $n_i=N_i/N_{tri}$, where $N_i$ is the number of such triads. With this definition, the number of positive links and the density of such links become $L_+=(3N_0+2N_1+N_2)/(N-2)$, and $\rho=L_+/L$, respectively, where $L=\binom{N}{2}$ is the total number of links and $L_{+}$ is the number of positive links in the system. In this respect, the positive link density and the system energy can be written as $\rho=n_0+2n_1/3+n_2/3$ and $U=n_1+\epsilon n_3$, respectively. At every time step, we flip a randomly chosen link with probability \cite{Shojaei2019}
\begin{equation}
p=\frac{1}{1+e^{\beta\Delta U(t)}}
\end{equation}
where $\beta$ can be considered as the inverse of the stochasticity in the individual behavior. Also, $\Delta U(t)$ represents the total energy change due to the link flipping in every time step $t$. This model resembles the Glauber dynamics used in simulations of kinetic Ising models at a given temperature $T=1/k\beta$ \cite{Glauber1963}. In fact, this provides a more pragmatic situation in which the tension in the system can either decrease or increase at any given time step, while for finite $\beta$ the tension decreases on average \cite{Shojaei2019}. Thus, the system can escape from jammed states, which are local minima in the energy landscape of the system \cite{marvel2009energy}. We investigate the dynamics of the above model for various initial configurations $\rho_0$ and energies $\epsilon$.

\begin{figure}[ht]
\begin{center}
\includegraphics[scale=.28]{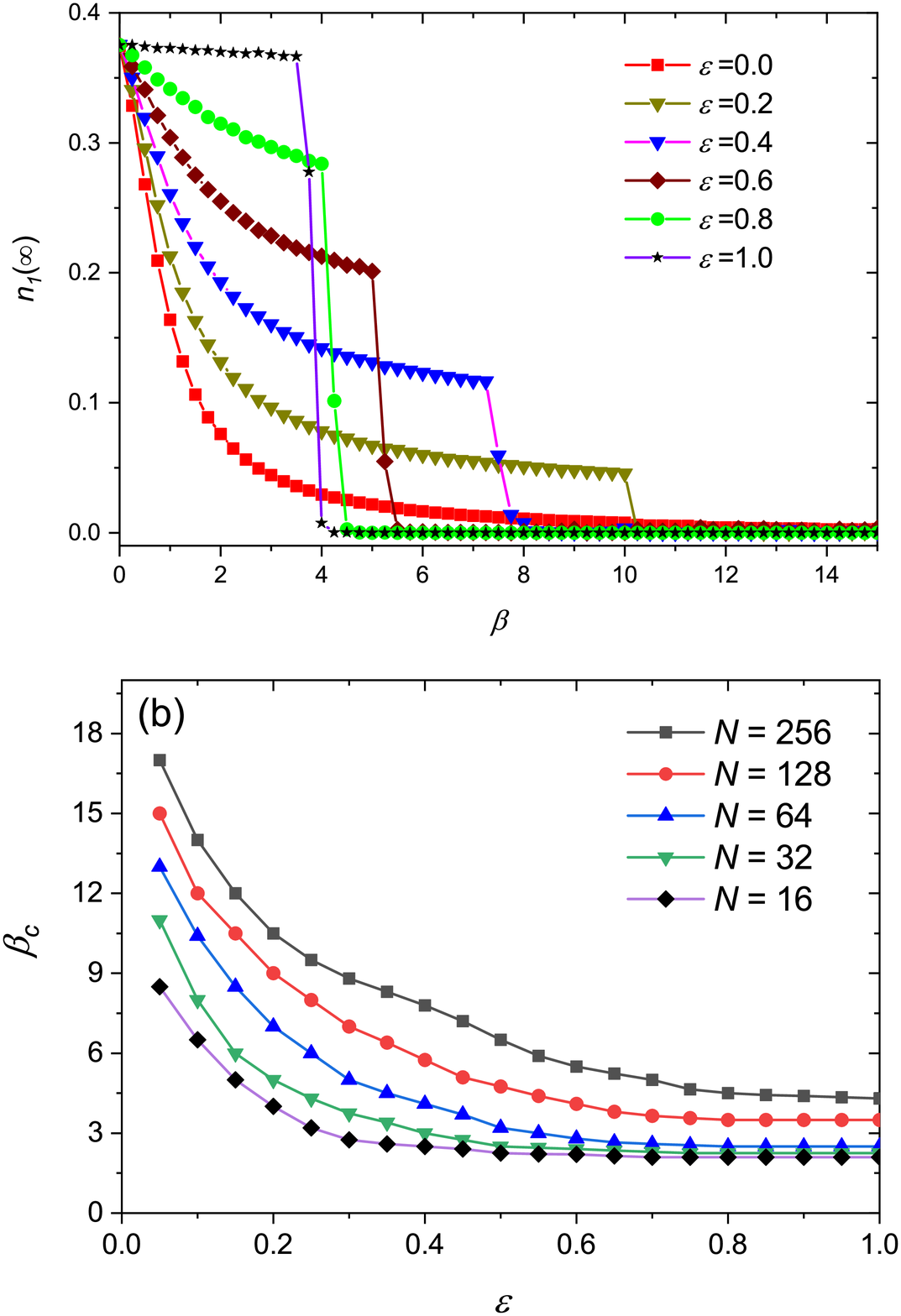}
\caption{(a) The $\beta$ dependency of large time behavior of final densities $n_1$, as the inverse of order parameter in the system, for $\rho_0 = 0.4$. Different energies of $\epsilon$ are shown with different symbols. The system transitions to an ordered state, at some values of $\beta=\beta_c$. The system size in all plots is $N=256$. (b) The critical values of $\beta_c$ versus $\epsilon$ and for different system sizes $N$. As can be seen, $\beta_c$ goes to infinity for large networks or small $\epsilon$.}
\label{fig1}
\end{center}
\end{figure}

\begin{figure}[ht]
\begin{center}
\includegraphics[scale=.28]{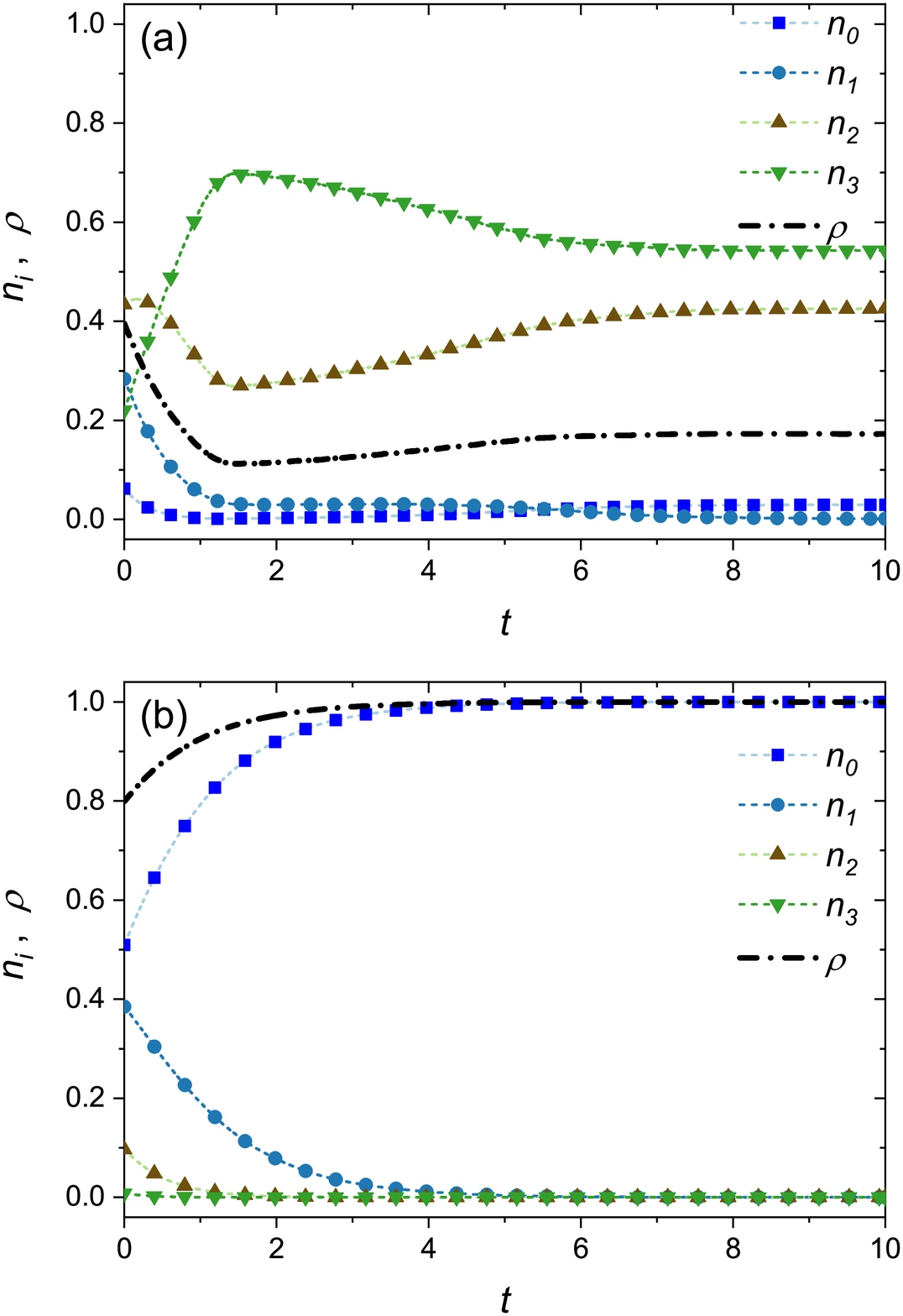}
\caption{The time evolution of triad densities, $n_i$, with $\epsilon=0.2$ and for initial positive link densities of (a) $\rho_0=0.4$ and (b) $\rho_0=0.8$. The triad density $n_1$ vanishes and thus the system approaches into a stable (weak) balanced state. The dot-dashed lines in both figures indicate the corresponding time evolution of positive link density $\rho$. For all plots, $\beta>\beta_c$ (see Fig.~\ref{fig1}) and the system size is $N=256$.}
\label{fig2}
\end{center}
\end{figure}

\begin{figure}[ht]
\begin{center}
\includegraphics[scale=.28]{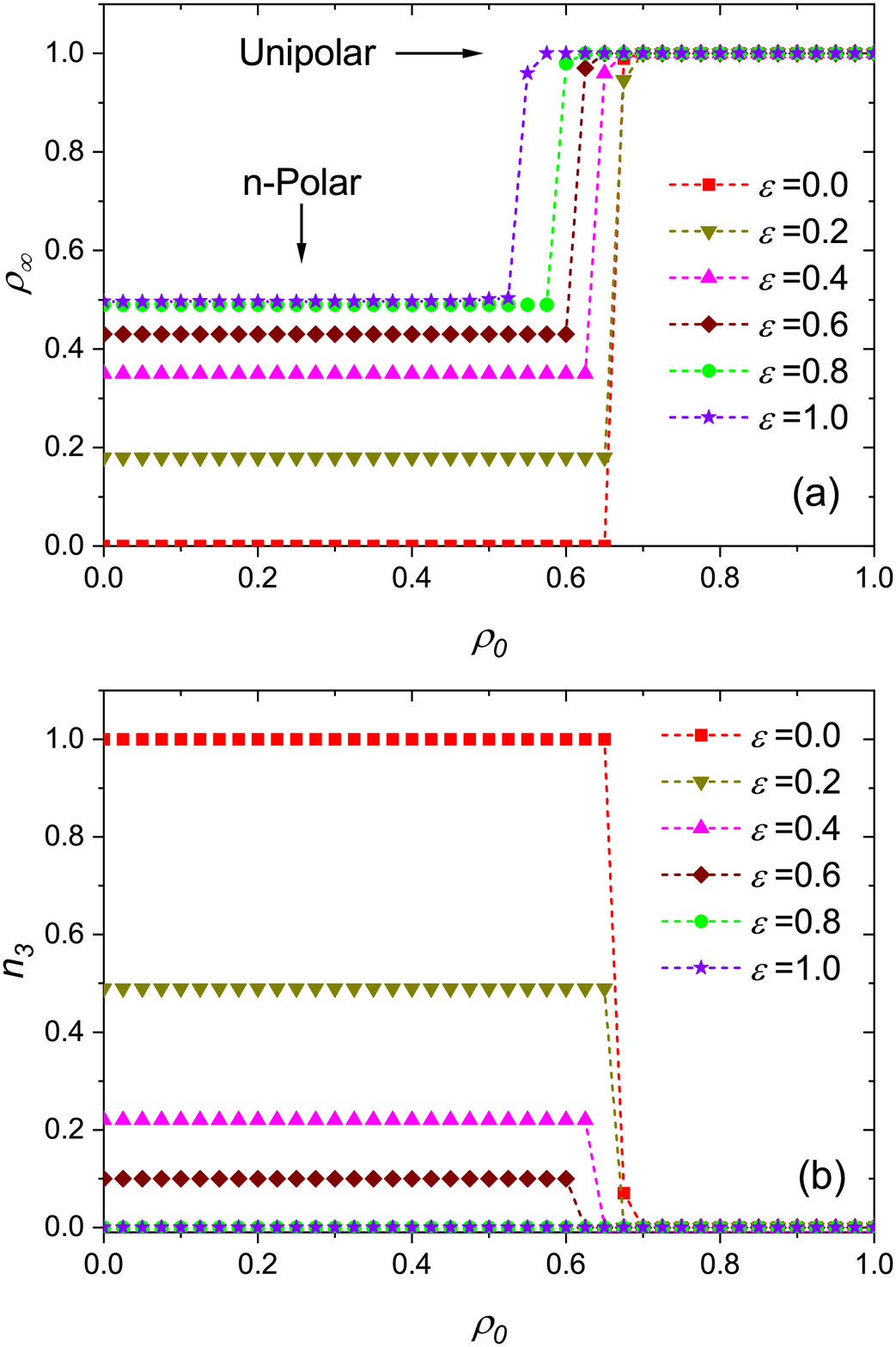}
\caption{(a) The final positive link density $\rho$ and (b) the final triad density $n_3$ versus $\rho_0$ for different $\epsilon$. As can be seen, the final fate of the system is a unipolar phase for any arbitrary $\epsilon$ when $\rho_0>\rho_0^c$. For $\rho_0\leq\rho_0^c$, $n$-polar states with $n\geq 2$ emerge. Other parameters are the same as in Fig.~\ref{fig2}.}
\label{fig3}
\end{center}
\end{figure}

\begin{figure}[ht]
\begin{center}
\includegraphics[scale=.28]{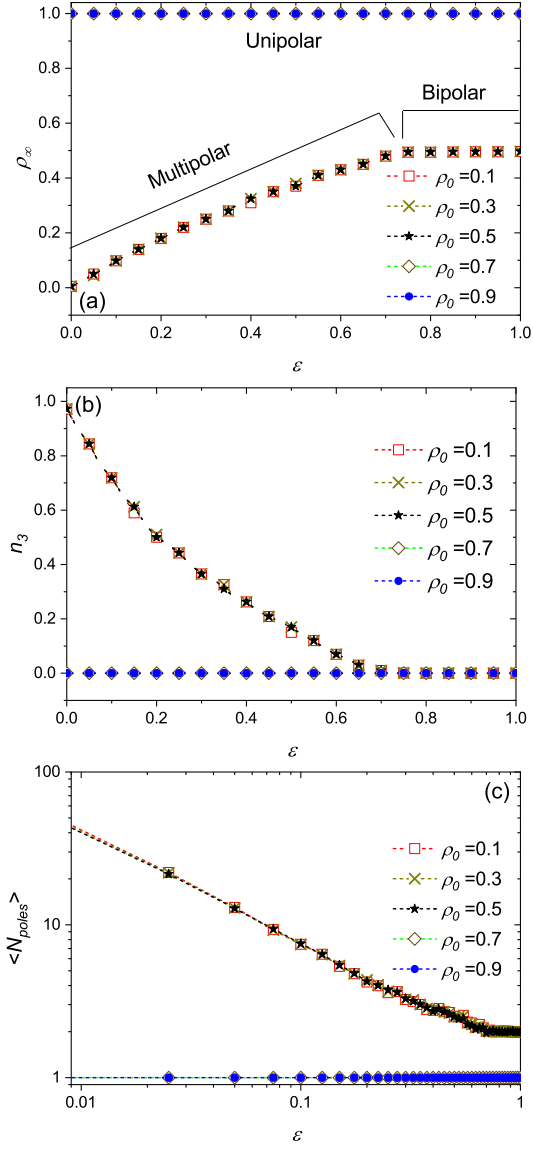}
\caption{(a) The final positive link density $\rho$ and (b) the final triad density $n_3$ versus $\epsilon$ for different $\rho_0$. A phase transition from bipolar states into multipolar ones occurs at $\epsilon_c\approx 0.67$ when $\rho_0\leq\rho_0^c$. (c) The log-log plot of the mean number of poles versus $\epsilon$ for different values of $\rho_0$. For $\rho_0\leq\rho_0^c$, the number of poles decreases as a power law form of $\sim\epsilon^{-0.8}$. Other parameters are the same as in Fig.~\ref{fig2}.}
\label{fig4}
\end{center}
\end{figure}

\begin{figure}[ht]
\begin{center}
\includegraphics[scale=.55]{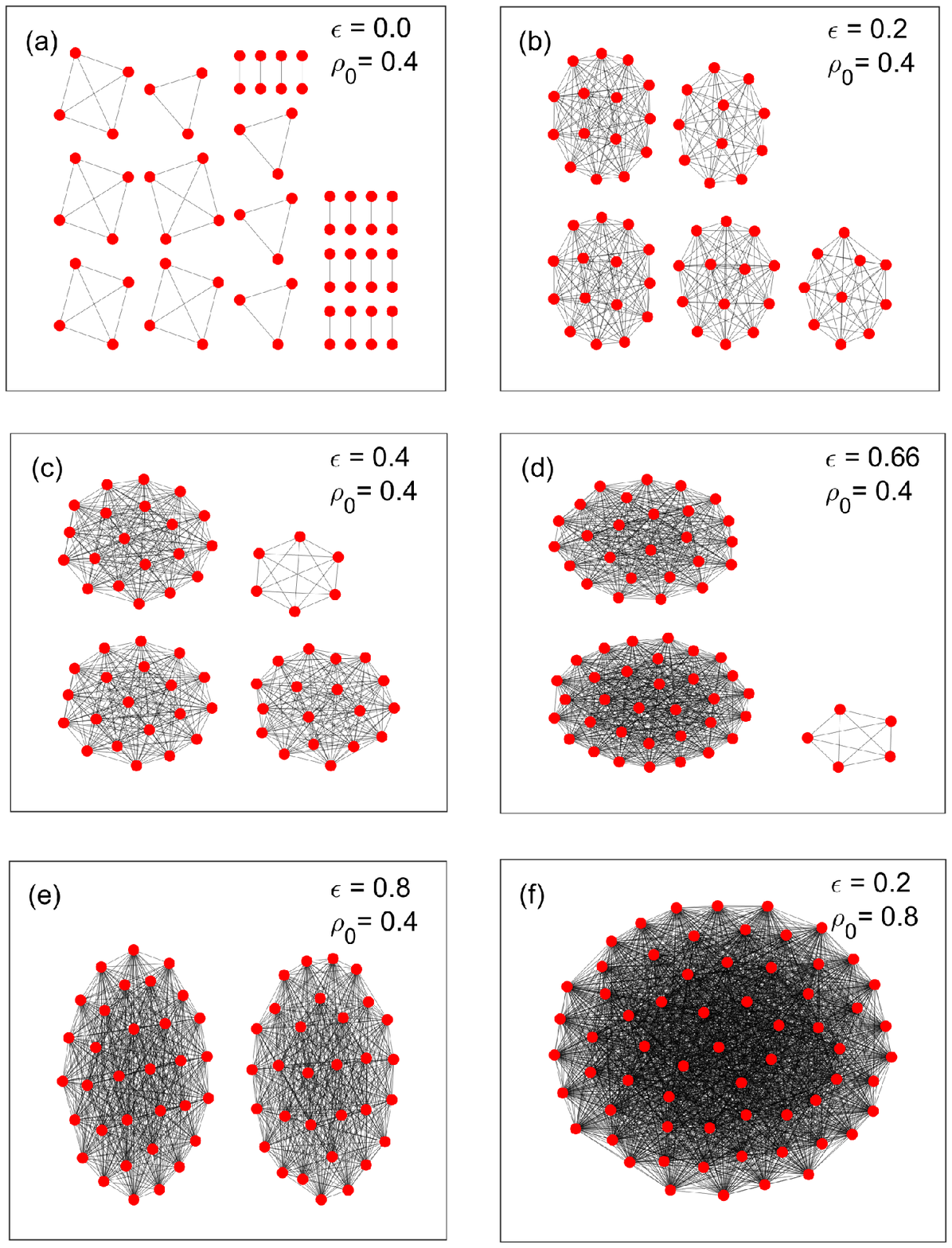}
\caption{Examples of six final states of the system: (a) to (d) represent multipolarity with different size of the poles. A bipolar and unipolar state are indicated in (e) and (f), respectively. Note that in (a) to (e) $\rho_0\leq\rho_0^c$ and in (f) $\rho_0>\rho_0^c$. The system size is $N=64$ for all graphs. Note that only friendly links are displayed.}
\label{fig5}
\end{center}
\end{figure}

\section{Numerical Results}
\label{diss}

Initially, we randomly distribute positive and negative links among all nodes so that the initial positive link density $\rho_0$ is obtained. Then, we start the dynamics by choosing an arbitrary link, randomly. To check the dependency of the final state of the system on the stochasticity in the individual's behavior, $\beta$, in Fig.~\ref{fig1}(a), we plotted the final values of triads of $n_1$ versus different $\beta$ and for different $\epsilon$. By taking into account the density $n_1$ as the inverse of order parameter (ordered state = a state without any unfavorable triadic relations, i.e., $n_1=0$), we find that for a given $\epsilon$, the system undergoes a phase transition from an unbalanced phase of $n_1 \neq 0$ into a stable weak balanced state with $n_1=0$ as $\beta$ crosses a critical value $\beta_c$ from bellow. As can be seen in Fig.~\ref{fig1}(b), this critical value $\beta_c$ is dependent on the value of the energy $\epsilon$. In fact, as $\epsilon$ decreases, $\beta_c$ increases. We note here that the value of $\beta_c$ is also dependent on the system size, and diverges for $N\to \infty$. This behavior is consistent with our previous work \cite{Shojaei2019,ManshourPRE_2021}, which can be considered as the special case of $\epsilon=1$ in the present work. This indicates that balanced states (weak or strong) are hardly reached in large systems as well as systems with $\epsilon\to 0$.

To show how the system evolves into a stationary (and stable) state, in Fig.~\ref{fig2}, we plot the dynamics of triad densities $n_0,n_1,n_2,n_3$ and positive link density $\rho$ for $\epsilon=0.2$ with initial conditions of $\rho_0=0.4$ and $0.8$ at $\beta>\beta_c$ (here $\beta=1.2\beta_c$). The system size here is $N=256$.  As can be seen, triads of type $\Delta_1$ disappear in all plots, i.e., $n_1(\infty)=0$. Thus, the final fate of the system can be three possible states due to the final values of other triad densities: unipolar ($n_2,n_3=0$), bipolar ($n_2\neq 0, n_3=0$) and mulipolar ($n_3\neq 0$). For example, the system approaches into a multipolar state in Fig.~\ref{fig2}(a) and a unipolar state emerges in Fig.~\ref{fig2}(b). Also, the dot-dashed lines in both plots represent the corresponding final positive link density $\rho$ for both initial densities of $\rho_0=0.4$ and $0.8$. To better understand the final states in the system, we present in Fig.~\ref{fig3}(a) the final positive link density $\rho_{\infty}$ versus $\rho_0$ for different values of $\epsilon$. As can be seen, if $\rho_0$ is greater than a critical value of $\rho_0^c$, the final phase is a unipolar state for all values of $\epsilon$. On the other hand, for $\rho_0\leq\rho_0^c$, multipolar ($n_3\neq 0$) and bipolar ($n_3=0$) states can emerge for small and large $\epsilon$, respectively as observed in Fig.~\ref{fig3}(b) which represents the final densities of $n_3$. We note here that this critical value of $\rho_0^c$ is dependent on $\epsilon$ and we will show later that it is indeed an unstable branch in the phase space of the system.

To better check $\epsilon$ dependency of the final state of the system, we also plot in Fig.~\ref{fig4}(a), the final density $\rho_\infty$ versus $\epsilon$ for different initial densities $\rho_0$. We find again that for $\rho_0$ above or bellow the critical value $\rho_0^c$, the system can reach a unipolar or a multi-bipolar state, respectively. For the case of $\rho_0\leq\rho_0^c$, the fate of the system can be either a bipolar or a multipolar state if the energy $\epsilon$ is larger or smaller than a critical value of $\epsilon^*\approx 0.67$, as observed in Fig.~\ref{fig4}(b). In fact, for $\epsilon\geq\epsilon^*$ the degree of tension associated to triads of type $\Delta_3$ is high enough that they cannot survive in the final state of the network. On the other hand, for $\epsilon<\epsilon^*$ we find that multipolar states with different sizes emerge. We are also interested in the properties of these emerging multipolar states. For example, Fig.~\ref{fig4}(c) demonstrates the mean number of poles $\left\langle N_{pole}\right\rangle$ for different values of initial densities $\rho_0$ and energies $\epsilon$, where $\left\langle ...\right\rangle$ represents an average over $500$ different realizations of the system. We see that when $\rho_0\leq\rho_0^c$, the mean number of poles decreases as a power law form $\left\langle N_{pole}\right\rangle\sim \epsilon^{-0.8}$ when $\epsilon<\epsilon^*$ and remains constant ($\left\langle N_{pole}\right\rangle=2$) if $\epsilon\geq \epsilon^*$. Also, for $\rho_0>\rho_0^c$, we have $\left\langle N_{pole}\right\rangle\to 1$ which indicates the unipolarity independent of $\epsilon$. It is noteworthy to mention here that the observed number of poles in real-world systems usually is not large and our results show that this can occur for a reasonable values of energies $\epsilon$ around $0.5$. Finally, in Fig.~\ref{fig5} we present six examples of possible network configurations corresponding to final states of the system for different values of $\epsilon$ and $\rho_0$. Indeed, Figs.~\ref{fig5}(a) to (d) represent four examples of multipolar states with different pole sizes and Fig.~\ref{fig5}(e) indicates a bipolar state. As we mentioned above, a unipolar state emerges for any values of $\epsilon$ when $\rho_0>\rho_0^c$ as indicated in Fig.~\ref{fig5}(f). Note that for all Fig.~\ref{fig5}(a) to (e), $\rho_0\leq\rho_0^c$ and for Fig.~\ref{fig5}(f) $\rho_0>\rho_0^c$.

\section{Mean-field approach}
\label{sec:analytic}

Since the system possesses large number of degrees of freedom, its exact time dependent dynamical equations are hard to obtain. In this respect, we search for a mean-field approximation for the rate equations, using the notations used in \cite{antal2005dynamics,Shojaei2019}. As we discussed before, it is appropriate to work with quantity $n_i$ which is the density of triads of type $\Delta_i$. Another useful quantity is the triad density $n_i^+$ ($n_i^-$) of type $\Delta_i$ that are connected to a positive (negative) link. $(3-i)N_i$ is the total number of positive links connected to triads of type $\Delta_i$, and thus the average number of such triads can be obtained as $N_i^+=(3-i)N_i/L_+$. Since, each link is connected to $N-2$ triads of any types, thus one can simply find that $n_i^+=N_i^+/(N-2)$. Similarly, we can write $n_i^-=N_i^-/(N-2)$ for a negative link. Consequently, we have
\begin{equation}
\begin{split}
n_i^+&=(3-i)n_i/(3n_0+2n_1+n_2) \\
n_i^-&=in_i/(n_1+2n_2+3n_3)
\end{split}
\label{n+-}
\end{equation}

By considering that $\rho$ is the probability of finding a positive link, the probability of flipping a \textit{positive} link is $\pi^+=p^+\rho$, with
\begin{equation}
p^+=\frac{1}{1+e^{\beta \Delta U_{+-}}}
\label{p_p}
\end{equation}
and of flipping a \textit{negative} link is $\pi^-=p^-(1-\rho)$, with
\begin{equation}
p^-=\frac{1}{1+e^{\beta \Delta U_{-+}}}
\label{p_p}
\end{equation}
where $\Delta U_{+-}$ and $\Delta U_{-+}$ are the energy difference due to the flipping a positive and a negative link, respectively. In fact, the transition probabilities $p^+$ and $p^-$ are the two pivotal parameters that drive the system dynamics. 

For an each update at step $j$, we have
\begin{equation}
L_+(j+1)-L_+(j)=-\pi^+ + \pi^-
\end{equation}
Since each time step equals $L$ updates, the rate equation for (average) $\rho$ can be written as
\begin{equation}
\frac{d\rho}{dt}=-\pi^+ + \pi^-
\label{rho_rate}
\end{equation}

The energy difference due to the flipping of a positive and negative link in each update step equals to $(N_0^+-N_1^++\epsilon N_2^+)/N_{tri}$ and $-(N_1^--N_2^-+\epsilon N_3^-)/N_{tri}$, respectively. Thus we obtain
\begin{equation}
\begin{split}
U(j+1)-U(j)=&\pi^+(N_0^+-N_1^+\\
            &+\epsilon N_2^+)/N_{tri}\\
            &-\pi^-(N_1^--N_2^-\\
            &+\epsilon N_3^-)/N_{tri}
\end{split}
\end{equation}
Therefore, we find the rate equation of the total energy as
\begin{equation}
\frac{dU}{dt}=\pi^+\Delta U_{+-} + \pi^-\Delta U_{-+}
\label{U_rate}
\end{equation}
where
\begin{equation}
\begin{split}
\Delta U_{+-}&=+3(n_0^+-n_1^++\epsilon n_2^+)\\
\Delta U_{-+}&=-3(n_1^--n_2^-+\epsilon n_3^-)
\end{split}
\label{Delt_u}
\end{equation}

Also, the rate equations for all triad densities, $n_i$, can be obtained in a similar way which are as follows:
\begin{equation}
\begin{split}
\frac{dn_0}{dt}&=-3\pi^+ n_0^+ + 3\pi^-n_1^- \\
\frac{dn_1}{dt}&=-3\pi^+ n_1^+ - 3\pi^-n_1^- +3\pi^+ n_0^++3\pi^-n_2^- \\
\frac{dn_2}{dt}&=-3\pi^+ n_2^+ - 3\pi^-n_2^- +3\pi^+ n_1^++3\pi^-n_3^- \\
\frac{dn_3}{dt}&=-3\pi^- n_3^- + 3\pi^+ n_2^+
\end{split}
\label{dens_rate}
\end{equation}

As we mentioned before, the system dynamics are governed by the two transition probabilities of $p^+$ and $p^-$. In this respect, $p^+=p^-$ means that the probability of transition of a positive link into a negative one is equal to the transition probability in the reverse direction. For example, if $\beta\to 0$, we have $p^+=p^-=1/2$, and one can simply find from Eq.~\ref{rho_rate} that $d\rho/dt=1/2-\rho$ which yields
\begin{equation}
\rho(t)=1/2+(\rho_0-1/2)e^{-t}
\end{equation}
This demonstrates that for large $t$, $\rho$ tends to $1/2$ as expected in such a fully random situation. However, to find an exact solution for finite $\beta$ is not straightforward, and we will present a qualitative explanation. At first, note that for a finite $\beta$, the relation $\Delta U_{+-}=\Delta U_{-+}$ satisfies the condition $p^+=p^-$. On the other hand, by assuming that the system remains uncorrelated during its early stages of the evolution, the triad densities become $n_0 = \rho^3$, $n_1 = 3\rho^2(1 - \rho)$, $n_2 = 3\rho(1 - \rho)^2$, and $n_3 = (1 - \rho)^3$. By substituting these values into Eqs.~\ref{n+-} and \ref{Delt_u}, we find
\begin{equation}
\Delta U_{+-}=-\Delta U_{-+}=+3\{(3+\epsilon)\rho^2 - (2+2\epsilon)\rho + \epsilon\}
\label{Delt_u_rand}
\end{equation}
This relation along with the previous relation of $\Delta U_{+-}=\Delta U_{-+}$, indicates that whenever the positive link density $\rho$ satisfies $\Delta U_{+-}=0$ (or $\Delta U_{-+}=0$), the condition $p^+=p^-$ is reached. Based on Eq.\ref{Delt_u_rand}, the two solutions of $\Delta U_{+-}=0$ can be obtained as

\begin{equation}
\begin{split}
\rho'&=((1+\epsilon)-\sqrt{1-\epsilon})/(3+\epsilon)\\
\rho''&=((1+\epsilon)+\sqrt{1-\epsilon})/(3+\epsilon)
\end{split}
\end{equation}

\begin{figure}[ht]
\begin{center}
\includegraphics[scale=.32]{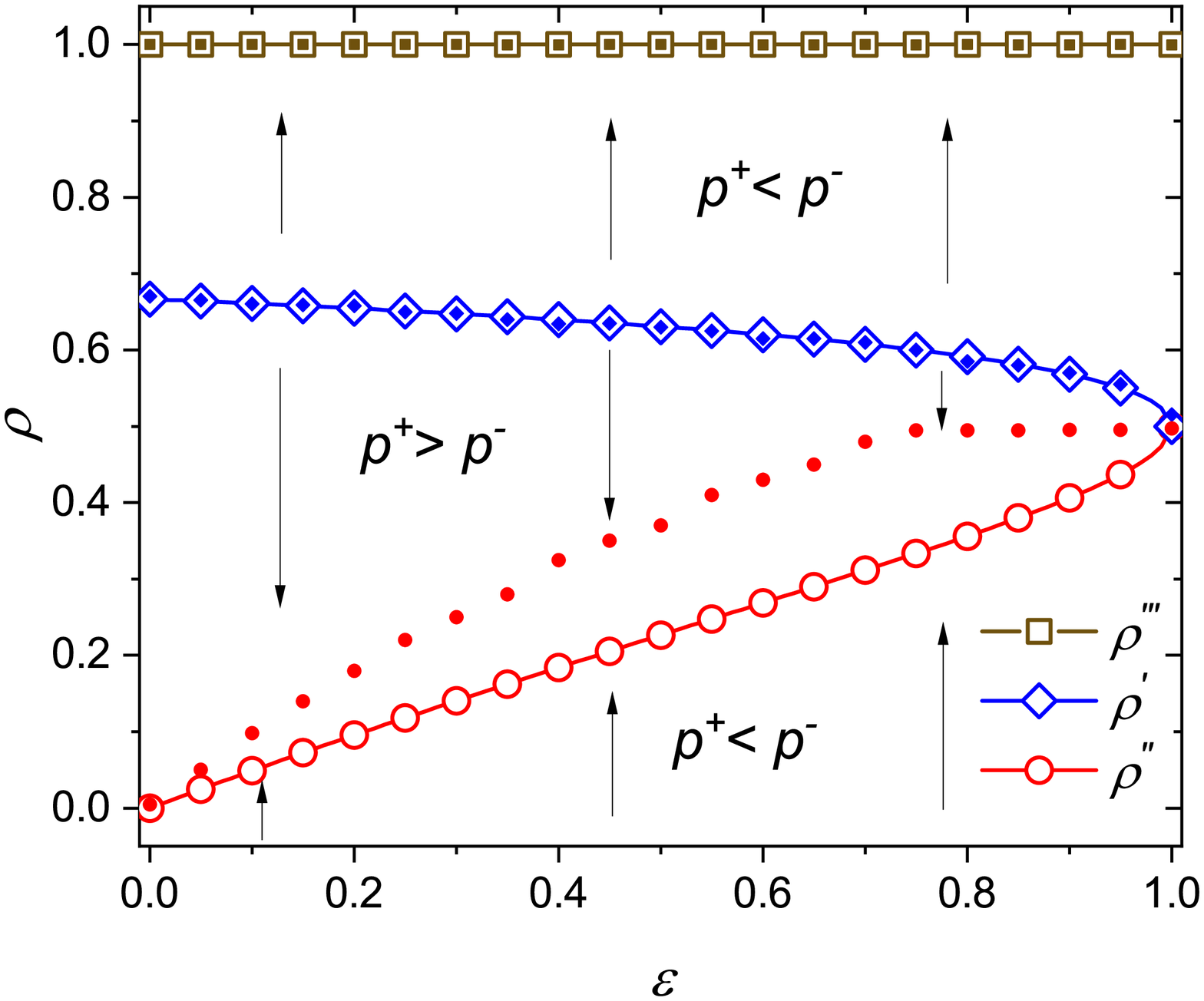}
\caption{Our analytical solutions obtained by using mean-field approximation. Blue unfilled diamonds, $\rho'$,  indicates the unstable solution, above and bellow which the system tends to another stable solutions of a unipolar ($\rho'''$) and a multi-bipolar ($\rho''$) state, respectively. The filled symbols represent our simulation results: Diamonds denote the values of phase transition points ($\rho_0^{c}$) obtained from Fig.~\ref{fig3}(a). Squares and Circles also represent final values of $\rho$ as indicated in Fig.~\ref{fig4}(a).}
\label{fig6}
\end{center}
\end{figure}

We plotted in Fig.~\ref{fig6}, these two solutions for different values of $\epsilon$. Indeed, for $\rho<\rho''$, we have $\Delta U_{+-}>0$ (or $\Delta U_{-+}<0$) which means that $p^+<p^-$. This indicates that negative links will be flipped into positive ones with higher probability, which on average increases $\rho$ until it reaches to $\rho''$ where $p^+=p^-$. By similar mechanism, if $\rho>\rho''$ then $p^+>p^-$, i.e., positive links will change to negative ones with higher probability, and this decreases $\rho$ until it again reaches $\rho''$. However, $\rho=\rho'$ is an unstable solution, since for $\rho>\rho'$, we have $p^+<p^-$ which increases the number of positive links until $\rho$ reaches its maximum value $\rho=\rho'''=1$, where $\rho'''$ is the stable unipolar state. For $\rho<\rho'$, we have $p^+>p^-$ which decreases the number of positive links until $\rho=\rho''$. Briefly, our mean-field analysis shows that a bifurcation occurs for $\epsilon<1$, with a stable branch, $\rho''$, for $\rho\leq 1/2$ and an unstable branch, $\rho'$, for $\rho>1/2$. Filled symbols in Fig.~\ref{fig6} represent our simulation results. In fact, filled diamonds represent values of transition points $\rho_0^c$ as observed in Fig.~\ref{fig3}(a). Filled squares and filled circles also show respectively two final possible states of unipolar and $n$-polar phases with $n\geq2$ represented in Fig.~\ref{fig3}(a) and Fig.~\ref{fig4}(a). This demonstrates that our mean-field approximation is mostly in agreement with our numerical simulations, and can well explain the phase space behavior of the system.

\section{Conclusion}
\label{conc} In social balance theory, triads with various
interactions are typically grouped into balanced and unbalanced
states. Such binary identification may lead to a globally balanced
situation which are either uniform or bipolar. On the other hand,
many real world situations exhibit multipolarity which have gained
much less attention in the literature. In this work, we showed
how considering a triad which contains all negative links as less
unbalanced than a triad with only one negative link, can lead to
an eventual state which contain multipolar communities. Our
stochastic dynamics was chosen in accordance with Glauber dynamics
in the presence of randomness $\beta$. We described the transition
to the multipolar state as a function of $\epsilon$ and $\rho_0$
and showed various phase diagrams. The number of final poles
crucially depends on the value of $\epsilon$ and can grow very
large as $\epsilon$ is reduced considerably. We also provided a
mean field calculation which showed how decreasing $\epsilon$ from
its standard value leads to a bifurcation with a stable and an
unstable branch which was mostly consistent with our numerical
simulations. We observed that our model typically leads to
multipolar states with roughly homogenous pole size distribution.
An interesting question to investigate is the conditions under
which a heterogenous size distribution may emerge in a multi-polar
society.

\section*{Acknowledgment}
PM would like to gratefully acknowledge the Persian Gulf University Research Council for support of this work.
AM also acknowledges support from Shiraz University Research Council.


%

\end{document}